\documentclass[aps,prx,reprint,superscriptaddress]{revtex4-2}
\usepackage{graphicx}
\usepackage{array}
\usepackage{amssymb}
\usepackage{amsfonts}
\usepackage{amsmath}
\usepackage{mathrsfs}
\usepackage{color}
\usepackage{booktabs}
\usepackage{threeparttable}
\usepackage{multirow}
\usepackage{subfigure}
\usepackage{times}
\usepackage{epsfig}
\usepackage{chngpage}
\usepackage{latexsym}
\usepackage{mathrsfs}
\usepackage{bm}
\usepackage{caption}
\usepackage{hyperref}
\hypersetup{colorlinks=true, citecolor=blue, urlcolor=blue, linkcolor=blue}

\begin{document}
\captionsetup[figure]{labelfont={bf},name={Fig.},labelsep=space,justification=raggedright}

\title{Universal scaling relation and criticality in metabolism and growth of \textit{Escherichia coli}}

\author{Shaohua Guan}
\altaffiliation{S.G. and Z.Z. contributed equally to this work.}
\affiliation{Defense Innovation Institute, Chinese Academy of Military Science, Beijing 100071, China}
\affiliation{CAS Key Laboratory of Theoretical Physics, Institute of Theoretical Physics, Chinese Academy of Sciences, Beijing 100190, China}
\author{Zhichao Zhang}
\altaffiliation{S.G. and Z.Z. contributed equally to this work.}
\affiliation{School of  Electronic Engineering, North China University of Water Resources and Electric Power, Zhengzhou 450046, China}
\author{Zihan Zhang}
\affiliation{CAS Key Laboratory of Theoretical Physics, Institute of Theoretical Physics, Chinese Academy of Sciences, Beijing 100190, China}
\affiliation{School of Physical Sciences, University of Chinese Academy of Sciences, Beijing 100049, China}
\author{Hualin Shi}
\email[Corresponding author: ]{shihl@itp.ac.cn}
\affiliation{CAS Key Laboratory of Theoretical Physics, Institute of Theoretical Physics, Chinese Academy of Sciences, Beijing 100190, China}
\affiliation{School of Physical Sciences, University of Chinese Academy of Sciences, Beijing 100049, China}
\affiliation{Wenzhou Institute, University of Chinese Academy of Sciences, Wenzhou, Zhejiang 325001, China}

\begin{abstract}
The metabolic network plays a crucial role in regulating bacterial metabolism and growth, but it is subject to inherent molecular stochasticity. Previous studies have utilized flux balance analysis and the maximum entropy method to predict metabolic fluxes and growth rates, while the underlying principles governing bacterial metabolism and growth, especially the criticality hypothesis, remain unclear. In this study, we employ a maximum entropy approach to investigate the universality in various constraint-based metabolic networks of \textit{Escherichia coli}. Our findings reveal the existence of universal scaling relations across different nutritional environments and metabolic network models, similar to the universality observed in physics. By analyzing single-cell data, we confirm that metabolism of \textit{Escherichia coli} operates close to the state with maximum Fisher information, which serves as a signature of criticality. This critical state provides functional advantages such as high sensitivity and long-range correlation. Moreover, we demonstrate that a metabolic system operating at criticality takes a compromise solution between growth and adaptation, thereby serving as a survival strategy in fluctuating environments.
\end{abstract}

\maketitle

\section{Introduction}
Molecular stochasticity in metabolic processes is a crucial factor in cellular physiology, impacting both bacterial growth and physiological adaptation \cite{wehrens2018stochasticity,schreiber2016phenotypic}. With the significant developments of single-cell technologies, it is feasible to track protein expression, metabolic fluxes, cellular growth and division at the single-cell level \cite{taniguchi2010quantifying, kiviet2014stochasticity, thomas2018sources}. Universal patterns, such as scaling laws in protein expression \cite{salman2012universal,brenner2015universal}, cellular growth and division \cite{iyer2014scaling,kennard2016individuality}, were revealed independently of bacterial species and nutritional environments. However, the underlying general principle governing the metabolic stochasticity of bacterial physiology remains elusive.

The tantalizing hypothesis of criticality in living systems, wherein these systems operate near critical points, has been investigated in various biological systems, including neural systems \cite{tkavcik2013simplest, tkavcik2015thermodynamics}, natural swarms \cite{attanasi2014finite,crosato2018thermodynamics}, and gene network of morphogenesis \cite{krotov2014morphogenesis}. The critical state is a state between order and disorder, which confers evolutionary advantages that enable living systems to effectively adapt to fluctuating environmental conditions \cite{hidalgo2014information}. Signatures of criticality, such as long-range spatiotemporal correlations and heightened sensitivity to stimuli, also contribute to biological functions \cite{stoop2016auditory,de2017critical}, which may potentially lead to several general trade-offs such as robustness and accuracy \cite{kauffman2003random}, robustness and evolvability \cite{torres2012criticality}, and robustness and flexibility \cite{chialvo2010emergent}. In bacterial metabolism, the coordinated behavior of individual biomolecules is crucial for sustaining high growth rates and minimizing lag time during nutrient shifts, highlighting an inherent trade-off between growth rate and physiological adaptation \cite{basan2020universal}. Exploring the criticality in bacterial metabolism and growth will shed light on the underlying principles governing metabolic phenotypic heterogeneity, cell-to-cell growth rate fluctuations, and the growth-adaptation trade-off. Universality is a fundamental concept in critical phenomena. Different systems may exhibit universal properties that are often irrelevant to most details of the system, so it is also important to reveal universal laws in biological systems.

To explore the criticality in bacterial metabolism and growth, based on the stoichiometric modeling of metabolic networks \cite{orth2010flux, o2015using, kauffman2003advances}, we introduce a maximum entropy model for metabolic networks of \textit{Escherichia coli}. The maximum entropy distribution of metabolic networks sampled from the feasible space of metabolic fluxes has been used to explain and predict the growth rate fluctuations and metabolic fluxes \cite{de2016growth,de2018statistical,fernandez2019maximum,fernandez2020statistical}. However, the maximum entropy model of metabolic networks still lacks systematic studies on diverse metabolic network models. With the discovery of new genomic and biochemical knowledge, the metabolic networks are constantly updated and periodically released \cite{reed2003expand, feist2007genome, orth2011comprehensive}. Moreover, simple metabolic network models can be constructed depending on the metabolic pathways of interest \cite{orth2010reconstruction,hadicke2017ecolicore2}. It will be of great interest to study the universality and criticality in these metabolic networks of varying complexity. Furthermore, the criticality, occurring at the boundary of the order-disorder phase transition, is regulated by an effective parameter known as the Lagrangian multiplier in the maximum entropy model. Inferring this effective parameter from the single-cell data can provide insights into whether bacterial metabolism and growth processes operate in the vicinity of the critical state.

In this study, we simulate the maximum entropy distributions across various nutritional environments and metabolic networks, revealing universal scaling relations between growth rate and Fisher information. We compare the parameter derived from single-cell data with the parameter obtained through maximum Fisher information. Furthermore, we quantitatively discuss the trade-off between growth and adaptation. We focus on metabolic networks and experimental data from Escherichia coli, and our approach is applicable to studying the metabolism and growth of other bacteria.

\section{Results}
\subsection{Order parameter and Fisher information}

The maximum entropy distribution of metabolic network is determined by the average growth rate and the feasible solution space of metabolic fluxes \cite{de2016growth}, depending on the specific details of metabolic models and the nutritional environment, written as
\begin{equation}
p(\boldsymbol{v}|\beta)=\frac{1}{\mathcal{Z}(\beta)}e^{\beta^e \frac{\lambda(\boldsymbol{v})}{\lambda_{max}^e}}.
\end{equation}
$\beta^e$ as a dimensionless parameter is the product of the Lagrange multiplier $\beta$ and $\lambda_{max}^e$ which is the maximum growth rate under environment $e$. The Lagrange multiplier $\beta$ can be understood as the regulatory capacity of a bacterial regulatory network. In metabolic networks, higher growth rate states require accurate matching between metabolic fluxes, implying that more precise regulatory capacity (large $\beta$) is required to achieve higher growth rates (large $\lambda$).

We studied several common metabolic models of \textit{Escherichia coli}: the Core model \cite{orth2010reconstruction}, iJR904 \cite{reed2003expand}, iAF1260 \cite{feist2007genome}, and iJO1366 \cite{orth2011comprehensive}. Among these models, the Core model is a simplified carbon metabolic model, the others are genome-level metabolic models. The maximum entropy (MaxEnt) distribution of metabolic fluxes is obtained by the hit-and-run sampling method (see Appendix~\ref{dd}). As shown in Fig.~\ref{f1}(a), with the gradual increase of the dimensionless parameter $\beta^e$, the probability distribution of the normalized growth rate gradually shifts to the right. The width of the probability distribution is narrow when $\beta^e$ is extremely large or small, and wider when $\beta^e$ is moderate. The MaxEnt distributions in various nutritional environments and metabolic models have similar trends (see Supplementary Fig. 1).

\begin{figure}
\includegraphics[width=8.6 cm]{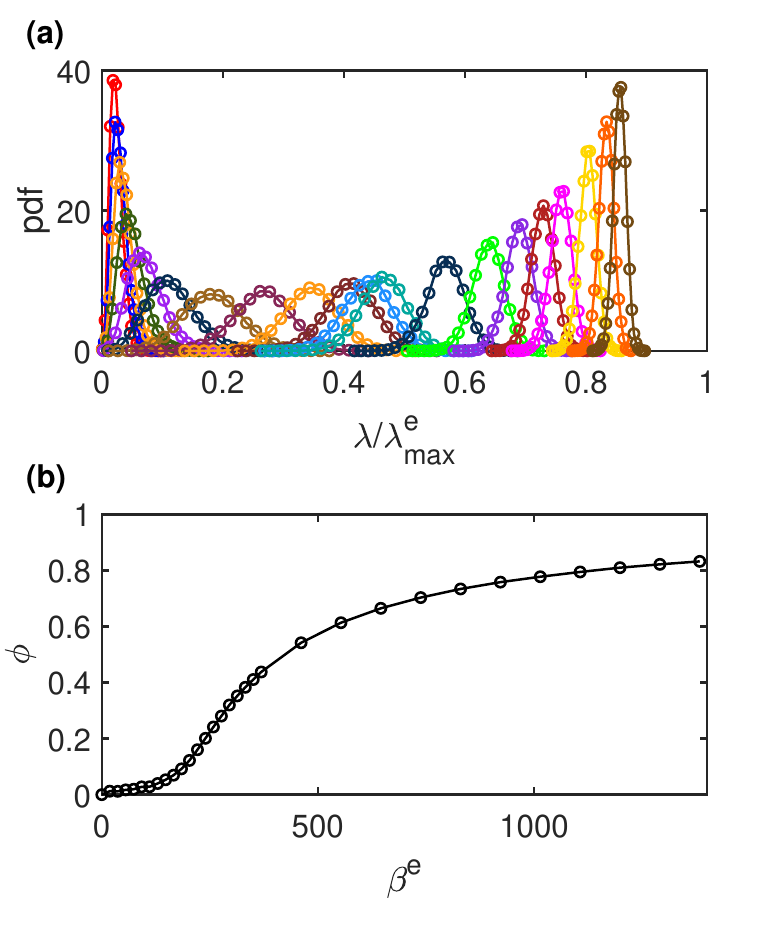}
\caption{{The probability distribution of normalized growth rate and order parameter change with $\beta^e$.} {(a)} The probability density functions of normalized growth rate under different $\beta^e$. The probability density distribution is calculated from the MaxEnt sampling data (iJR904 model), and it gradually moves to the right as $ \beta^e$ increases. The exchange flux of glucose in iJR904 is set to be -10 mmol/(gDW$\cdot$hour), which results in the maximum growth rate $\lambda_{max}^e = 0.9219\text{hour}^{-1}$. The range of $\beta$ is from $0$ to $1500$. (b) The order parameter $\phi$ defined in Eq.~(\ref{order_eq}) as a function of $\beta^e$. The order of MaxEnt distribution in iJR904 ranges from the disordered state ($\phi=0$) to the ordered state ($\phi=1$) by adjusting the dimensionless parameter $\beta^e$. The uniform distribution ($\beta^e = 0$) corresponds to the disordered state ($\phi=0$) and the maximum growth rate state ($\beta^e \to \infty$) corresponds to the ordered state ($\phi=1$).}
\label{f1}
\end{figure}
\begin{figure}
\includegraphics[width=8.6 cm]{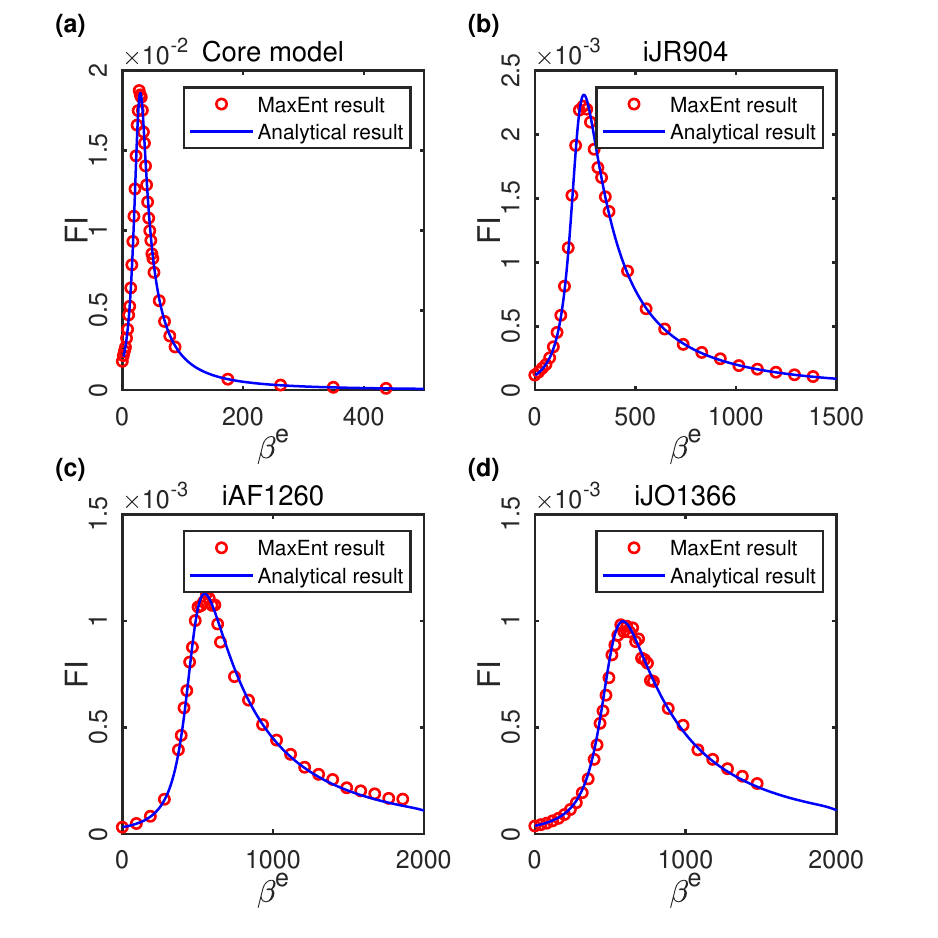}
\caption{{The relation between Fisher information (FI) and $\beta^e$ in different metabolic models.} The subfigures represent the relationship between Fisher information and $\beta^e$ in the Core model, iJR904 model, iAF1260 model, and iJO1366 model, respectively. Points in the subfigures are the sampling results of the MaxEnt distributions, and the lines stand for the fitted analytical results.}
\label{FI}
\end{figure}

In the metabolic models, we defined the order parameter as the degree between the uniform distribution state (disordered) and the maximum growth rate state (ordered), which is
\begin{equation}\label{order_eq}
\phi(\beta^e)= \frac{1}{N_{flux}}\sum_i \left|\frac{\overline{v_i}(\beta^e) - \overline{v_i}(\beta^e=0)}{\overline{v_i}(\beta^e \to \infty) - \overline{v_i}(\beta^e=0)} \right| .
\end{equation}
$\overline{v_i}(\beta^e)$ is the average of flux $v_i$ in the MaxEnt distribution, and $N_{flux}$ represents the number of fluxes in the metabolic model. The order parameter $\phi=0$ indicates that the metabolic system is in a disordered state, corresponding to the uniform distribution with a small average growth rate. As $\beta^e$ approaches infinity, the order parameter $\phi$ gradually increases to 1, corresponding to the flux balance analysis solution with maximum growth rate (see Fig.~\ref{f1}(b)).

In order to quantify the characteristics of the MaxEnt distributions in different models, we introduced Fisher information, which is an important quantity not only in information theory but also in statistical physics, e.g., in the geometric representation of thermodynamics, where Fisher information (FI) is a geometric metric \cite{janke2004information}. In the MaxEnt model, Fisher information is defined as
\begin{equation}\label{fffiii}
F = \frac{\partial \overline{\lambda}(\beta^e)/\lambda_{max}^e}{\partial \beta^e} = \overline{ (\lambda(\beta^e)/\lambda_{max}^e - \overline{\lambda}(\beta^e)/\lambda_{max}^e)^2 }.
\end{equation}
It can be seen that the Fisher information represents the degree of the normalized average growth rate changing with parameter $\beta^e$, and is also the variance of the normalized growth rate. In Fig.~{\ref{FI}}, Fisher information increases with $\beta^e$ and then decreases after reaching the maximum value $FI^{\ast}$ at $\beta^{e\ast}$, which depends on the specific metabolic model. Comparing with Fig.~{\ref{f1}}(b), the metabolic system is found to be ordered for $\beta^e>\beta^{e\ast}$ and disordered for $\beta^e<\beta^{e\ast}$.

The MaxEnt distribution is determined by the parameter $\beta^e$ and the uniform distribution under $\beta^e=0$. As shown in \cite{de2016growth}, the beta distribution has a strong fitting ability to the uniform distribution of normalized growth rate $p(\lambda/\lambda_{max}^e)|_{\beta^e=0}$, which is the state density function of the normalized growth rate. We used the beta distribution to fit the uniform distributions of normalized growth rate in different metabolic models, and the relationships of $\overline{\lambda}/\lambda_{max}^e \mathit{\sim} \beta^e$ and $FI \mathit{\sim} \beta^e$ are obtained and consistent with the simulation results (see Appendix~\ref{ee}).

\subsection{Universal scaling relations}
\begin{figure}
\includegraphics[width=8.6 cm]{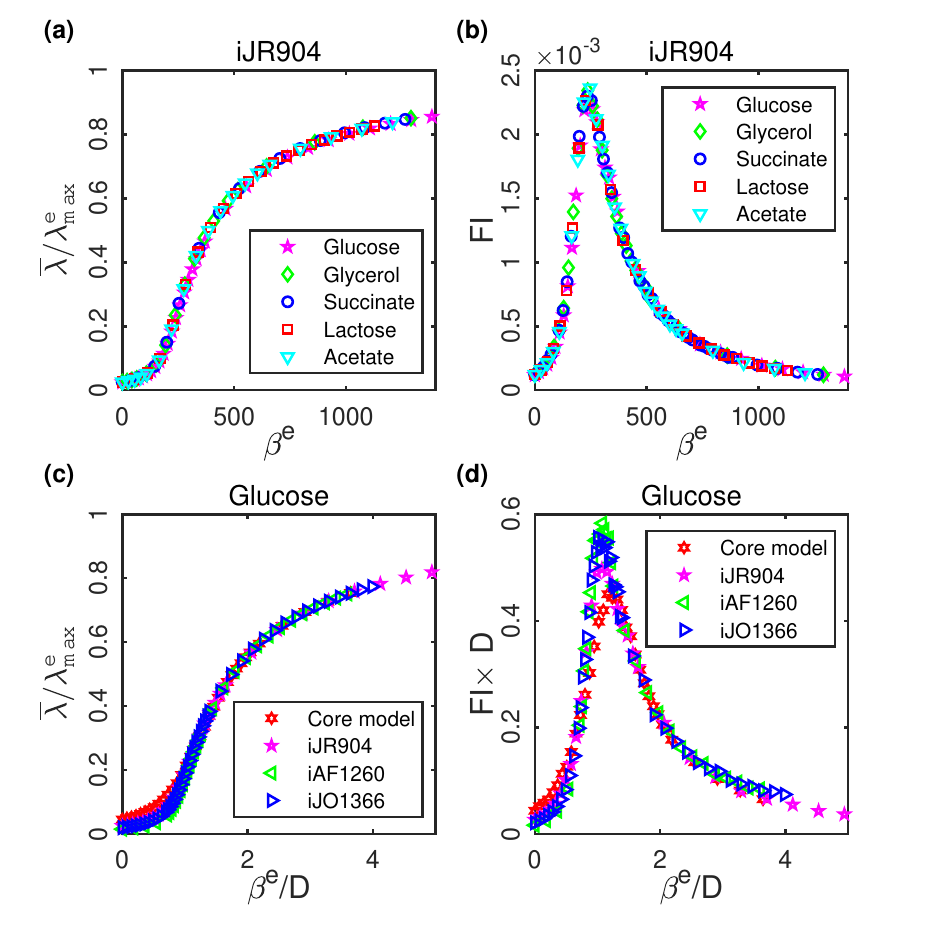}
\caption{{Universal relations of $\overline{\lambda}/\lambda_{max}^e \mathit{\sim} \beta^e$ and $FI \mathit{\sim} \beta^e$.} The simulated data of iJR904 model collapse together under five kinds of carbon resources, showing universal relations of $\overline{\lambda}/\lambda_{max}^e \mathit{\sim} \beta^e$ ({a}) and $FI \mathit{\sim} \beta^e$ ({b}). In the same nutritional environment, after scaled by dimension $D$, the simulated data of the Core model, iJR904, iAF1260, and iJO1366 also collapse together, showing universal relations of $\overline{\lambda}/\lambda_{max}^e \mathit{\sim} \beta^e/D$ ({c}) and $FI\times D \mathit{\sim} \beta^e/D$ ({d}).}
\label{scaling}
\end{figure}

The nutritional environment and metabolic network structure determine the feasible space $\mathcal{P}$ of metabolic fluxes, resulting in significant differences in the choice of metabolic pathways under different environments. For example, after the nutritional shift-up, metabolic pathways need to be switched to maintain efficient cellular growth and metabolism. Although the feasible spaces of diverse environments and models are quite different, the MaxEnt models of metabolic networks still exhibit universal laws. As shown in Fig.~\ref{scaling}(a), the relationships of $\overline{\lambda}/\lambda_{max}^e \mathit{\sim} \beta^e$ under four different carbon sources (Glucose, Lactose, Succinate and Glycerine) in iJR904 model collapse together, suggesting that the transition of carbon sources has little effect on the normalized average growth rate $\overline{\lambda}/\lambda_{max}^e$. As shown in Eq.~(\ref{fffiii}), Fisher information is the slope of the relationship of $\overline{\lambda}/\lambda_{max}^e \mathit{\sim} \beta^e$. According to the universal scaling relation in different nutritional environments, the relationships between $FI$ and $\beta^e$ also collapse together  (Fig.~\ref{scaling}(b)). The normalized average growth rate and Fisher information in these universal relations are similar to the spontaneous magnetization and susceptibility in magnetic systems, which exhibit universal curves among various metals \cite{pathria2016statistical}. The difference is that temperature is scaled by the critical temperatures of different metals, and $\beta$ is scaled by the maximum growth rate under different nutritional environments.

The dimensionality $D$ of the metabolic model (see Appendix~\ref{cc}), analogous to the scale of physical system, represents the degrees of freedom of the metabolic network model and has a large impact on the shape of feasible space $\mathcal{P}$. Therefore, even in the same nutritional environment, the normalized growth rate distributions in different models have large differences. However, by scaling the dimensionless parameter $\beta^e$ with the model dimension $D$, the relationships of $\overline{\lambda}/\lambda_{max}^e \mathit{\sim} \beta^e/D$ in four different metabolic models collapse on a universal relation (Fig.~\ref{scaling}(c)). The relationships of $FI\times D \mathit{\sim} \beta^e/D$ also collapse together (Fig.~\ref{scaling}(d)), showing finite-scale scaling for metabolic networks of different scales. It is analogous to the finite-size scaling law in natural swarms \cite{attanasi2014finite,holubec2021finite}. The system scale of swarms is easy to adjust, however, the dimensionality of metabolic network is difficult to change due to the limited number of metabolic network models of \textit{Escherichia coli}. Interestingly, the parameter with the maximum Fisher information (MaxFI) $\beta^e / D \approx 1$ is consistent with the critical point $k_c \approx 1$ of maximum entropy models in other biological systems \cite{mora2010maximum,mora2011biological,tkavcik2015thermodynamics}. It indicates that there is a universal critical point in the maximum entropy model of biological systems, independent of the specific details of biological systems. The mechanism of this universal feature in biological criticality deserves further study.

\subsection{Experimental test of criticality with single-cell data}
The signature of criticality is a divergence of Fisher information in the thermodynamic limit, and a maximum of Fisher information for the finite-size systems \cite{prokopenko2011relating}. An interesting question is whether the distributions of growth rate measured in biological experiments are in the vicinity of the critical state. We used previously published data \cite{kennard2016individuality}, where \textit{Escherichia coli} were stably grown in stable conditions for long periods of time, to test the criticality hypothesis in the context of metabolism and growth. The key point is whether the parameter $\beta^e$ of experimental growth rate distribution is close to the $\beta^e$ in the MaxFI state.

\begin{figure}
\includegraphics[width=8.6 cm, trim= 10 70 10 0]{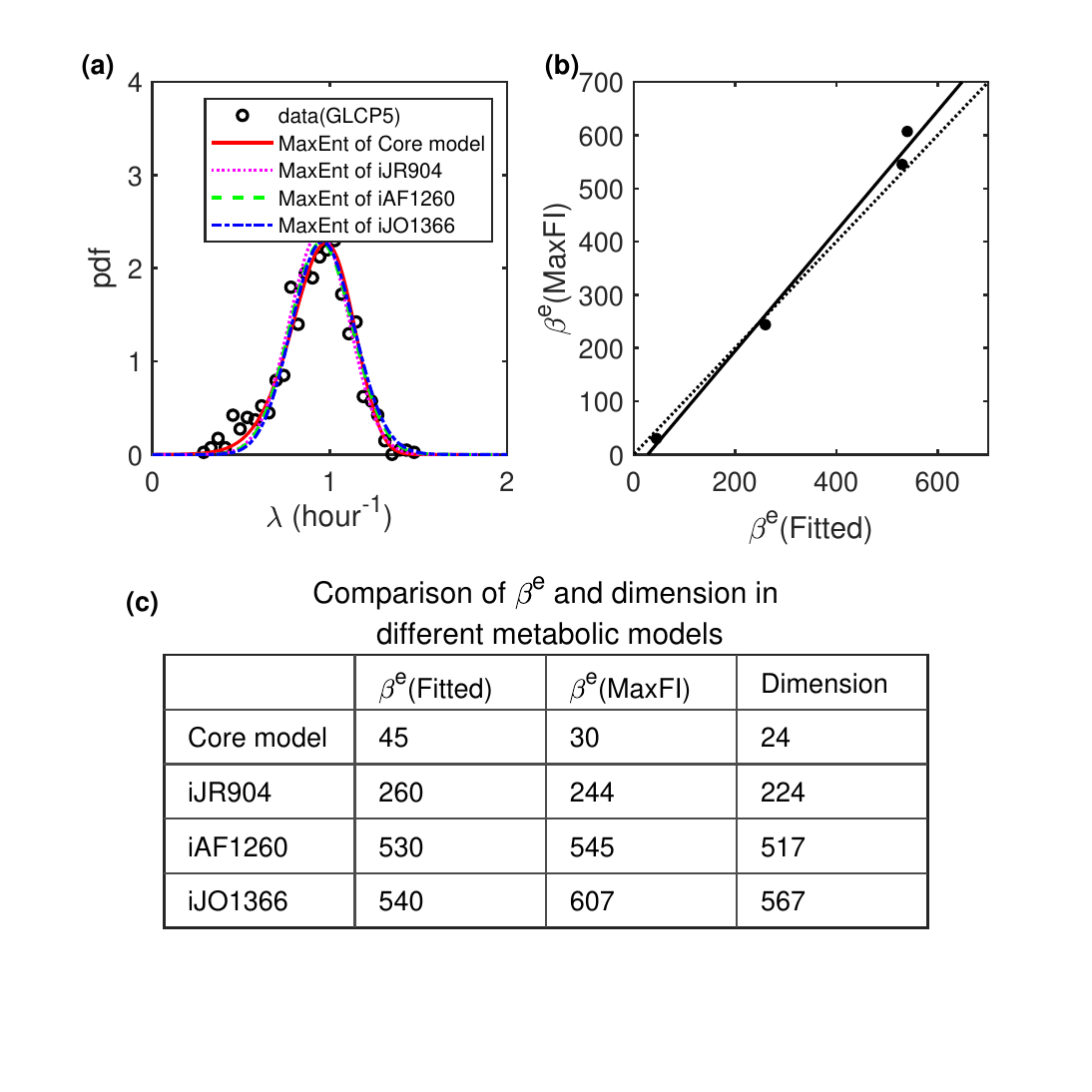}
\caption{{Comparison of measured growth rate distribution and MaxEnt distributions of different models.} {(a)} Fitting the MaxEnt distributions of different metabolic models to the probability density function of the single-cell growth rate measured in the experiment \cite{kennard2016individuality}. The circles are the single-cell experimental data, and the lines are the fitted MaxEnt distributions of metabolic models. {(b)} In different metabolic models, the best-fitted parameters $\beta^e$ are close to the $\beta^e$ with MaxFI, showing a linear relationship. The four points represent four metabolic models which are listed in {(c)}. The solid line is the best-fitted linear relation between $\beta^e(\text{Fitted})$ and $\beta^e(\text{MaxFI})$. The dotted line is used to indicate the case where $\beta^e(\text{Fitted})$ and $\beta^e(\text{MaxFI})$ are equal. {(c)} The $\beta^e(\text{Fitted})$, $\beta^e(\text{MaxFI})$ and Dimensions of four metabolic network models for \textit{Escherichia coli}.}
\label{fit}
\end{figure}

By fitting the parameters $\beta$ and $\lambda^e_{max}$, the MaxEnt distributions of growth rate in four different models and the experimental growth rate distribution have an excellent match (Fig.~\ref{fit}(a)). The best-fitted $\beta^e$ vary widely across different models, but is close to the $\beta^e$ in the MaxFI state, showing a linear relationship between the best-fitted $\beta^e$ of single-cell data and the parameter $\beta^e$ with the MaxFI (Fig.~\ref{fit}(b)). It suggests that the Fisher information of the experimental growth rate distribution of \textit{Escherichia coli} is close to the MaxFI independent of the choice of metabolic models.

According to the definition of Fisher information in Eq.~(\ref{fffiii}), the normalized average growth rate is most sensitive to the parameter $\beta^e$ when the Fisher information is maximum. In the MaxEnt model of metabolic networks, the Lagrange multiplier $\beta$ represents the regulation capability of bacterial regulatory network \cite{de2016growth}, and the parameter $\lambda_{max}^e$ corresponds to the external nutritional environment. Therefore, when Fisher information is maximized, the normalized average growth rate is highly sensitive to the variation of internal regulation and external environment, which is conducive to the rapid adjustment of growth rate to internal and external fluctuations, and thus has an evolutionary advantage.

\subsection{Long-range correlation of metabolic fluxes near the MaxFI state}
The large correlation between individuals is an important feature of criticality, which may provide living systems with coordinated behavior across space and time. We will show that the correlations between metabolic fluxes and growth rate exhibits a clear structure of strong (anti-)correlation near the MaxFI state. 

\begin{figure*}
\includegraphics[width=12 cm, trim= 150 20 120 0]{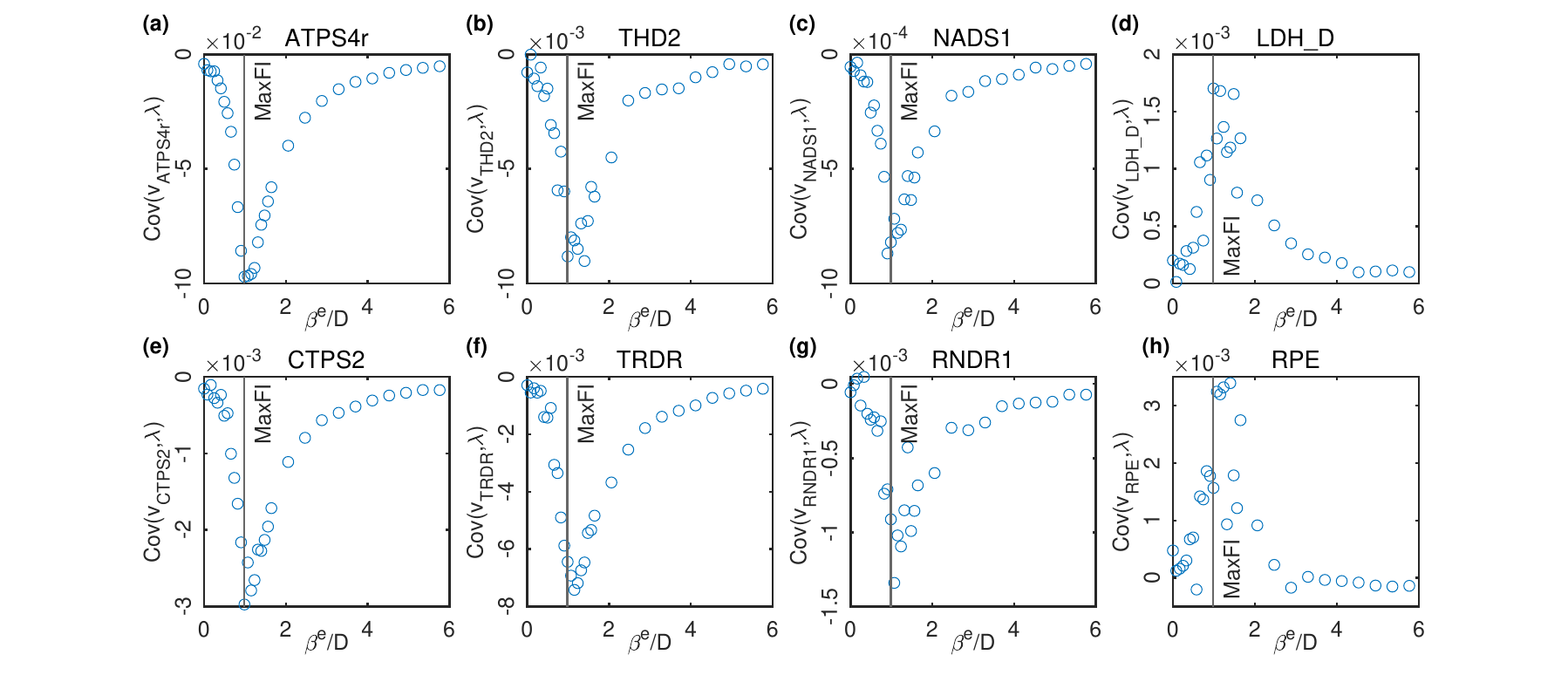}%
\caption{{The correlations between metabolic fluxes and growth rate $Cov(v_i, \lambda)$ changing with $\beta^e/D$.} Under different $\beta^e$, metabolic fluxes in (a)-(c) and (e)-(g) are negatively correlated with growth rate and metabolic fluxes in (d) and (h) are positively correlated with growth rate. When $\beta^e/D$ is close to the value with MaxFI, the $Cov(v_i,\lambda)$ is near the maximum or minimum value.}
\label{corre}
\end{figure*}

Since bacteria live in a constantly changing environment, the metabolic system needs to rapidly adjust metabolic flux states in response to environmental fluctuation. Considering the large scale of the metabolic network, weak associations between metabolic fluxes are not conducive to information transfer in the metabolic system. Therefore, the long-range correlation between metabolic fluxes is crucial for the response speed of the metabolic system. The correlation between metabolic flux and growth rate ($Cov(v_i, \lambda)$) can be calculated through the MaxEnt sampling results and written as
\begin{equation}\label{equ_cov}
C(v_i,\lambda)|_{\beta^e}=\frac{1}{N_{s}}\sum_j(v_i^j|_{\beta^e} - \overline{v_i}(\beta^e))(\lambda^j|_{\beta^e} - \overline{\lambda}(\beta^e)).
\end{equation}
$v_i^j$ denotes the $j$th sample of $i$th flux, and $N_s$ is the number of samples in the MaxEnt sampling. In Fig.~\ref{corre}, the correlations between metabolic fluxes and growth rate have maximum (or minimum) values near the MaxFI state as $\beta^e$ changes. Most of the metabolic fluxes in the metabolic network do not have a direct relationship with growth rate. Fluctuation of one metabolic flux usually first affects the adjacent metabolites and metabolic fluxes in the metabolic network. After information transfer through multiple metabolites and fluxes, It can indirectly affect the growth rate, similar to the long-range correlation in physical systems with short-range interaction. The universal feature of the relationships between flux correlation and $\beta^e$ in Fig.~\ref{corre} suggests that in the MaxFI state, large correlations between metabolic fluxes lead to long-range correlation in the metabolic network. Variations of metabolic fluxes caused by internal cellular noise or external environmental fluctuations can significantly affect the metabolic state and growth rate, which facilitates information transfer in metabolic networks.

\subsection{The trade-off between growth and adaptation}
In the MaxEnt model, the large $\beta^e$ corresponds to the fast-growing phenotypes that favor bacterial survival in a specific environment. However, phenotypic heterogeneity as a general survival strategy can provide high adaptation for biological populations in fluctuating environments \cite{acar2008stochastic,ackermann2015functional,schreiber2016phenotypic,patange2018escherichia},  suggesting a fundamental trade-off between growth and environmental adaptability \cite{basan2020universal}. The competition between stable growth and flexible adaptation is reflected in physical systems as a competition between energy and entropy. But on the contrary physical systems tend to be in a state of low energy and bacteria tend to grow fast. In order to discuss this fundamental trade-off in the MaxEnt models of metabolic networks, we quantified the environmental adaptability as the number of metabolic states of the MaxEnt distribution in a given $\beta^e$, since a large number of metabolic states provide potential capacity to speed up the response to nutritional shifts. We will demonstrate that the MaxFI state of the metabolic system employs a compromise solution to deal with the competition between growth and adaptation.

\begin{figure}
\includegraphics[width=8.6 cm]{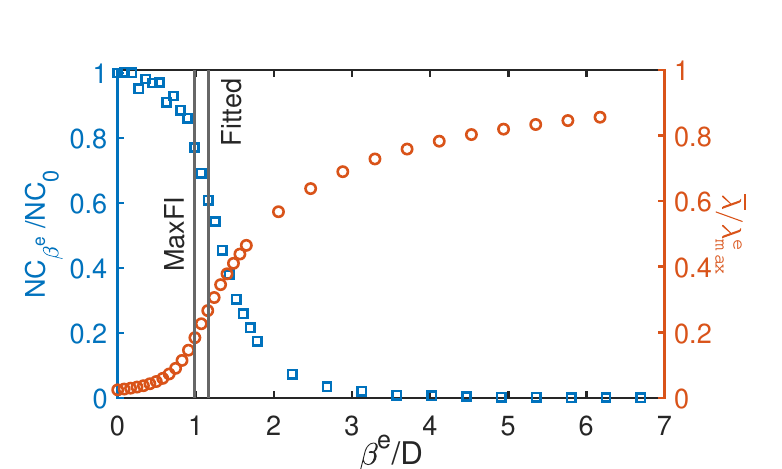}
\caption{{The normalized cluster number $NC_{\beta^e}/NC_0$ of metabolic states and the normalized average growth rate $\overline{\lambda}/\lambda^e_{max}$ as functions of $\beta^{e}/D$.} The cluster number of metabolic states shrinks as $\beta^e/D$ increases. As $\beta^e/D$ tends to infinity, the cluster number converges to the unique metabolic state at which the growth rate is maximized. Compared to the normalized cluster number, the relationship between $\overline{\lambda}/\lambda_{max}^e$ and $\beta^e/D$ has an opposite trend. The intersection of these relations has a moderate growth rate and cluster number of metabolic states. The $\beta^e(\text{Fitted})$ and $\beta^e(\text{MaxFI})$ in iJR904 (solid lines) are close to the intersection.}
\label{af1}
\end{figure}

Although the metabolic fluxes could in principle be in any state of the feasible space $\mathcal{P}$ regardless of $\beta^e$, only a small fraction of the metabolic flux states can be taken due to the finite sampling number or finite bacterial cells, especially the metabolic state is unique in the maximum growth rate state. We introduced a clustering algorithm to cluster the neighboring MaxEnt sampling points in $\mathcal{P}$ as a metabolic state  (see Appendix~\ref{ff}). As shown in Fig.~\ref{af1}, the normalized cluster number decreases with $\beta^e$, and the normalized average growth rate increases with $\beta^e$, showing a trade-off between fast growth and environmental adaptability. A large number of clusters means that the metabolic state has greater uncertainty, that is, the MaxEnt distribution of the metabolic network has a larger entropy. It suggests that the growth-adaptation trade-off is a manifestation of the universal energy–entropy competition in bacterial metabolism and growth. The intersection of growth rate and cluster numbers in Fig.~\ref{af1} represents the balance of energy–entropy competition, and $\beta^e/D$ at the intersection is close to the $\beta^e/D$ with MaxFI and the fitted $\beta^e/D$ of single-cell data, indicating that the metabolic system adopts a compromise between faster growth and stronger environmental adaptability.

In summary, we illustrated the fundamental trade-off between growth and adaptation in the MaxEnt model of metabolic networks. A compromise strategy between these two optimization objectives can be accomplished by the criticality, which has the MaxFI and long-range correlation between metabolic fluxes. It suggests that operating near criticality could be a survival strategy for bacteria in fluctuating environments.

\section{Discussion}
In this work, we revealed universal scaling relations and several signatures of criticality in metabolism and growth of \textit{Escherichia coli} from the maximum entropy approach. The universal scaling relations between growth rate and $\beta^e$, as well as the scaling relations between Fisher information and $\beta^e$, emerge independently of the metabolic network models and nutritional environments. By comparing the single-cell data with the maximum Fisher information state, it suggested that the metabolism and growth of \textit{Escherichia coli} may operate near the critical state. We found several functional advantages at criticality numerically, such as high sensitivity to stimuli, the long-range correlation between fluxes, and a compromise strategy between fast growth and environmental adaptability.

The universal scaling relations suggest the inherent scale invariance of metabolic networks. The scaling laws of the cell-size and division-time distributions emerge by scaling their average values \cite{iyer2014scaling}. However, we scaled the parameter $\beta$ by a theoretical quantity $\lambda_{max}^e$ and network dimension $D$, providing a broader view to discover new scaling laws. Criticality as a candidate general principle has been widely explored in biological systems. We provided a metabolic version of biological criticality which is the boundary between two distinct phases or regimes: a disordered one with many metabolic states and an ordered one with a fast growth rate. However, due to the limited number of effective metabolic models, it is still difficult to systematically study the finite-size scaling of metabolic networks.

The stochasticity of metabolism and growth is inevitable. Noise in gene expression, signal transduction, and metabolism affects the processes of bacterial growth and division, but also provides a mechanism for bacterial adaptation. We show that for bacterial metabolism and growth, the system is in a critical state with several functional advantages. Thus, it indicates that stochasticity in metabolic processes is delicately regulated by biological systems through gene regulatory networks and protein interaction networks. This general idea will contribute to revealing the design principles of biological networks.

The constraints of flux balancing construct a feasible solution space of metabolic fluxes, which can be analogized to the state space of physical systems. The maximum entropy distribution over the solution space is constrained by the average growth rate, corresponding to the energy function in the canonical ensemble. However, the growth rate is not always a valid objective function, just as the energy function needs to be transformed into other quantities under different ensembles. For example, the goal of metabolic systems in multicellular organisms may be to maintain the specific biological function, rather than to continue to grow and divide. As another example, for the energy metabolism network, maximizing the rate of energy metabolism is a more appropriate goal. Therefore, in the maximum entropy model of the metabolic network, it is necessary to select the appropriate flux as the objective function according to the specific situation. Recent studies argued that thermodynamics and ensemble theory emerge as the limiting behaviors of statistics \cite{lu2022emergence,qian2022internal}. This generalized ensemble theory might be able to provide a theoretical basis of maximum entropy model of biological systems. 

In the traditional flux balance analysis, the influence of protein in the metabolic process is not considered. The concentration of enzymes is constrained by molecular crowding \cite{beg2007intracellular}. For a metabolic process, a larger metabolic flux requires more enzymes to catalyze the metabolic process, and more protein expression has an impact on the growth of bacteria \cite{mori2016constrained}. For example, after adding protein constraints, it can correctly predict the transition of metabolic pathways between aerobic and anaerobic conditions, denoted as overflow phenomenon \cite{mori2019yield}. Adding more reasonable constraints will change the feasible solution space of metabolic fluxes, so it will be interesting to study the maximum entropy distributions in the solution space under more constraints in the future.

\begin{acknowledgments}
This work was supported by the Strategic Priority Program of Chinese Academy of Sciences (No. XDA17010504, XDPB15), the National Natural Science Foundation of China (Grant Nos. 12274426 and 12047503). S.G. and H.S. conceived the study and developed the theory. S.G. and Zhichao Zhang performed the simulations. S.G., Zhichao Zhang, and Zihan Zhang carried out the data analysis. All authors discussed the findings, drew the conclusions and wrote the paper. Shaohua Guan and Zhichao Zhang contributed equally to this work. The authors declare no competing interests.
\end{acknowledgments}

\appendix
\section{{\label{aa}}THE FEASIBLE SOLUTION SPACE $\mathcal{P}$ OF METABOLIC FLUXES}
The metabolic network uses a set of reactions to describe the metabolic state. With the huge development of biochemistry and biological technology, metabolic networks can be built at the genome level, which is usually termed as genome-scale metabolic model. 

Flux balance analysis is a method of analyzing the genome-scale metabolic model, which is combined with two main assumptions to avoid the amount of dynamic parameters. Assuming that the metabolic reactions in cells are at steady state, there are constraints on the reaction rates, written as
\begin{equation}
\dot{c}_{\mu}=\sum_{i}S_{i\mu}v_i =0.
\label{eq1}
\end{equation}
Here, $S_{i\mu}$ represents the stoichiometry of metabolite $\mu$ in reaction $i$, $c_\mu$ is the concentration of metabolite $\mu$, and $v_i$ is the rate of reaction $i$.

Making another assumption that cells would adapt reaction rates to maximize their growth rate, which is 
\begin{align}
max &:  v_{biomass} \label{eq2}\\
subject\;to &: \sum_{i}S_{i\mu}v_i=0 \label{eq3}.
\end{align}
The $v_{biomass}$ is an artificial reaction in the metabolic model, which equals to the growth rate $
\lambda$.

Considering that the concentrations of metabolites and enzymes limit their reaction rates, we used $v_i^u$ and $v_i^l$ to represent the bounds of the reaction rate $v_i$. It can be written as
\begin{equation}
\quad v_i^l<v_i<v_i^u .\label{eq4}
\end{equation}
If the reaction rate is a negative value, it means the reaction direction is reversed.

We used an efficient computational method to quickly perform flux variability analysis \cite{chan2018accelerating}. The result of flux variability analysis will be taken into Eq.~(\ref{eq4}) as the upper and lower bounds of the fluxes. 

In addition to the boundary conditions, there are also thermodynamic constraints that need to be considered \cite{qian2005thermodynamics}. Thermodynamically infeasible cycles satisfy the mass balance but violate the second law of thermodynamics. The maximum entropy flux distributions generated by our sampling method are inconsistent with the second law of thermodynamics. However, since we are mainly concerned with the distribution of growth rates, it does not significantly affect our results.

The flux balance analysis is the integration of  Eq.~(\ref{eq2}), Eq.~(\ref{eq3}) and Eq.~(\ref{eq4}), written as
\begin{align}
max &:  v_{biomass} \label{eq5}\\
subject\;to &: \sum_{i}S_{i\mu}v_i=0 \label{eq6}\\
&\quad v_i^l<v_i<v_i^u \label{eq7}.
\end{align}
Eq.~(\ref{eq6}) and Eq.~(\ref{eq7}) construct the feasible solution space $\mathcal{P}$ of the metabolic fluxes, while maximizing the growth rate in Eq.~(\ref{eq5}) makes the solution of the metabolic fluxes concentrated in the solution space where the growth rate is the largest.

\section{{\label{bb}}MAXIMUM ENTROPY MODEL OF METABOLIC NETWORK}
De Martino \textit{et al.} \cite{de2016growth} proposed a maximum entropy model (MaxEnt model) which has a Boltzmann-form distribution over the solution space constrained by metabolic flux balancing, written as
\begin{equation}
p(\boldsymbol{v}|\beta)=\frac{1}{\mathcal{Z}(\beta)}e^{\beta \lambda(\boldsymbol{v})}.
\end{equation}
Here, $\boldsymbol{v}$ refers to a possible fluxes state in $\mathcal{P}$ and $\lambda(\boldsymbol{v})$ is the growth rate of state $\boldsymbol{v}$. $\mathcal{Z}$ is defined as $\int_{\mathcal{P}} \exp(\beta \lambda(\boldsymbol{v})) d\boldsymbol{v}$. This Boltzmann-form distribution is the maximum entropy distribution constrained by the normalization $\int p(\boldsymbol{v}) d\boldsymbol{v} = 1$ and the average growth rate
\begin{equation}
\int p(\boldsymbol{v}|\beta)  \lambda(\boldsymbol{v}) d\boldsymbol{v}= \overline{\lambda}(\beta)=\lambda_{data},
\end{equation}
where $\lambda_{data}$ stands for the average growth rate obtained by measuring the cell population and $\beta$ is the Lagrange multiplier in the maximum entropy model. The growth rate distribution of the MaxEnt model has two asymptotic limit forms. $\beta\longrightarrow 0$ corresponds to the uniform distribution, and $\beta\longrightarrow \infty$ corresponds to the maximum growth rate state which is the solution of traditional flux balance analysis.

A transformation of the MaxEnt model is
\begin{equation}
p(\boldsymbol{v}|\beta)=\frac{1}{\mathcal{Z}(\beta)}e^{\beta\lambda_{max}^e \frac{\lambda(\boldsymbol{v})}{\lambda_{max}^e}}=\frac{1}{\mathcal{Z}(\beta)}e^{\beta^e \frac{\lambda(\boldsymbol{v})}{\lambda_{max}^e}}.
\end{equation}
$\lambda_{max}^e$ represents the maximum growth rate under the environment $e$. $\beta\lambda_{max}^e$ is denoted as $\beta^e$ which is a dimensionless parameter.

\section{{\label{cc}}METABOLIC NETWORK MODELS}
In this study, we employ four common metabolic network models of \textit{Escherichia coli}. The Core model~\cite{orth2010reconstruction} is reconstructed by the selected reactions, including 54 metabolites, 95 chemical reactions, and 20 exchange reactions. The iJR904 model~\cite{reed2003expand} is based on the annotation of the K-12 MG1655 genome updated in 2000. It is reconstructed by 931 reactions, containing 904 gene products. The iAF1260 model~\cite{feist2007genome} is the next iteration of metabolic network, reconstructed by updated genome annotation. It contains 1260 genes, 2077 metabolic reactions and 1039 unique metabolites. The iJO1366~\cite{orth2011comprehensive} is an update of iAF1260. It contains 1366 genes, 2251 metabolic reactions and 1136 unique metabolites. 

In a metabolic network, the stoichiometry matrix usually does not have full rank. The rank of the stoichiometry matrix is the dimension $D$ of the polytope $\mathcal{P}$. In these models, the dimensions are 24 (the Core model), 224 (iJR904), 517 (iAF1260), 567 (iJO1366).

The results of flux balance analysis and maximum entropy distributions are closely related to boundary conditions. In these four metabolic models, we select different carbon sources. When one carbon source is selected, there will be no other carbon sources. The uptake rates of different carbon sources are chosen as glucose$=-10$mmol/(gDW$\cdot$hour), glycerol$=-20$mmol/(gDW$\cdot$hour),  succinate=$-20$mmol/(gDW$\cdot$hour), lactose=$-15$mmol/(gDW$\cdot$hour), acetate=$-60$mmol/(gDW$\cdot$hour). The aerobic environment is selected as the simulation condition, but the aerobic or anaerobic environment does not affect the results of the paper. The other boundary conditions are the default values of the models. It should be pointed out that although boundary conditions affect the size and shape of the solution space, the relations between growth rate and Fisher information are universal after scaling.

\section{{\label{dd}}SAMPLING METHOD}
We used the package COBRA Toolbox V.3.0~\cite{heirendt2019creation}, which includes the sampling algorithm (the coordinate hit-and-run with rounding (CHRR) method~\cite{haraldsdottir2017chrr}), to do MaxEnt sampling. The solution space is an anisotropic polytope. Before sampling, the solution space was rounded by a maximum volume ellipsoid algorithm~\cite{zhang2003numerical} to avoid the impact of the flattening of the solution space on the sampling efficiency. The ellipsoid algorithm is putting a high-dimensional ellipsoid in and out of the solution space. By continuously adjusting the shape of the inner and outer ellipsoids, two ellipsoids that are very close to the solution space are finally obtained, and the sampling in one of the two ellipsoids has nearly the same distribution compared to sampling in the solution space. After doing the ellipsoid algorithm on the solution space, we used the hit-and-run algorithm to sample in the ellipsoid solution space, which is defined as follows:

1. Randomly select an initial point $\bm{x_1}$ in the high-dimensional ellipsoid solution space;

2. Choose a direction $\bm{\theta_1}$ in the high-dimensional ellipsoid solution space;

3. Draw a straight line along the direction $\bm{\theta_1}$ through the point $\bm{x_1}$, and intersect the ellipsoidized solution space at two points $\lambda_{max}^1$ and $\lambda_{min}^1$.

4. Generate a random number $\lambda_1$ between $\lambda_{max}^1$ and $\lambda_{min}^1$, then calculate the new point $\bm{x_2}=\bm{x_1}+\lambda_1\bm{\theta_1}$.

5. Repeat step 2 until enough sampling points were obtained.

The details of uniform sampling method are discussed in \cite{de2015uniform} and \cite{haraldsdottir2017chrr}. By modifying the probability of uniform sampling in step 4, that is, setting the probability of the next point to be consistent with the probability in the MaxEnt model by accepting and rejecting sampling, the MaxEnt distribution is finally obtained.

Other algorithms, such as the gaussian analytical approximation algorithm \cite{braunstein2017analytic}, provide faster computation speeds for large-scale metabolic networks, paving the way for the future systematic study of the finite-size scaling law of metabolic networks.

\section{{\label{ee}}FITTING THE MAXENT DISTRIBUTIONS}
In the maximum entropy model, $\beta=0$ is corresponding to the uniform distribution. For the uniform distribution obtained by sampling, the beta distribution can be used to fit the uniform distribution of normalized growth rate, written as
\begin{equation}
\rho(\lambda^e) = \frac{1}{\Phi(a,b)} {(\lambda^e)}^b(1-{\lambda^e})^a.
\end{equation}
${\lambda^e}$ represents $\lambda/\lambda_{max}^e$ and $\Phi(a,b)$ is the normalized function. This beta distribution contains two parameters $a$ and $b$ which can be obtained by fitting the uniform distribution of normalized growth rate. Starting from the uniform distribution, the normalized function $\mathcal{Z}(\beta^e)$ in the maximum entropy model can be written as
\begin{equation}
\mathcal{Z}(\beta^e)=\int_0^{1} \frac{1}{\Phi(a,b)} {(\lambda^e)}^b(1-{\lambda^e})^a e^{\beta^e\lambda^e} d\lambda^e.
\end{equation}
The distribution of normalized growth rate ${\lambda^e}$ is 
\begin{equation}
p(\lambda^e|\beta^e)=\frac{1}{\Phi(a,b)\mathcal{Z}(\beta^e)}{(\lambda^e)}^b(1-{\lambda^e})^a e^{\beta^e\lambda^e}.
\end{equation}
Then the normalized average growth rate in the maximum entropy model can be written as
\begin{equation}
\overline{\lambda^e}({\beta^e})=\frac{\partial ln\mathcal{Z}(\beta^e)}{\partial {\beta^e}}.
\end{equation}
After fitting the parameters of the beta distribution in uniform sampling, the relationship between ${\lambda^e}$ and $\beta^e$ is calculated, and then the relationship between Fisher information and $\beta^e$ in different models can be obtained by Eq.~(\ref{fffiii}). It can be seen in Fig.~\ref{FI} that the Fisher information obtained from uniform distribution fitting is consistent with the result obtained by sampling.\\

\section{{\label{ff}}CALCULATING THE NUMBER OF METABOLIC STATES}
Density-Based Spatial Clustering of Applications with Noise (DBSCAN) is applied to estimate the metabolic states of the sampling points. It is a density-based clustering algorithm \cite{ester1996density}. DBSCAN has two key parameters: eps and MinPts. The parameter eps is used to define the eps-neighborhood of a point $p$, a set of points whose distance to point $p$ is less than eps. If a point $q$ is in the eps-neighborhood of the point $p$, and the number of points in the eps-neighborhood of the point $p$ is more than or equal to MinPts, then the point $p$ is a core point and the point $q$ is directly density-reachable from point $p$. If there is a chain of points $p_1$, $p_2$,...,$p_n$ and $p_i$ is directly density-reachable from $p_{i-1}$, the point $p_n$ is density-reachable from $p_1$. If a point $p$ and a point $q$ are density-reachable from another point $g$, the point $p$ is density-connected to the point $q$. If the point $p$ belongs to the cluster $\mathbb{C}$, all the points which are density-reachable from $p$ also belong to the cluster $\mathbb{C}$. Any two points in the same cluster are density-connected to each other. That is the neighborhoods of any point in the cluster $\mathbb{C}$ are also in the cluster $\mathbb{C}$. Applying DBSCAN to the sampling data, the number of clusters is used to stand for the number of metabolic states.

Considering that the different fluxes vary widely and the sampling points are high-dimension data, we preprocessed the sampling data by scaling and PCR. The fluxes whose variations among all sampling points under different $\beta$ are less than $10^{-4}$ are ignored. Other fluxes in all sampling points under different $\beta$ are scaled to a range of 1 to 10. The scaled sampling data is obtained. Based on the scaled sampling data under $\beta = 0$, ten principal components are selected by the Principal Component Analysis \cite{martin1979multivariate}. It is implemented by the Stats package of the R language. Other scaled sampling data under different $\beta$ are reduced to 10 dimensions by the same PCR method. Thus, a group of 10-dimension datasets is obtained from the sampling data under different $\beta$ values. The 10-dimensional dataset for each $\beta$ value were analyzed by DBSCAN. The DBSCAN was done under eps = 6 and MinPts =1 by the dbscan function in fpc package of R language.

%

\end{document}